# A Framework for Fairer Machine Learning in Organizations

Lily Morse[1], Mike Horia Mihail Teodorescu[2,3], Yazeed Awwad[3], Gerald Kane[2]

[1] John Chambers College of Business and Economics, West Virginia University [2] Carroll School of Management, Boston College; [3] D-Lab, Massachusetts Institute of Technology.

## ABSTRACT

With the increase in adoption of machine learning tools by organizations risks of unfairness abound, especially when human decision processes in outcomes of socio-economic importance such as hiring, housing, lending, and admissions are automated. We reveal sources of unfair machine learning, review fairness criteria, and provide a framework which, if implemented, would enable an organization to both avoid implementing an unfair machine learning model, but also to avoid the common situation that as an algorithm learns with more data it can become unfair over time. Issues of behavioral ethics in machine learning implementations by organizations have not been thoroughly addressed in the literature, because many of the necessary concepts are dispersed across three literatures – ethics, machine learning, and management. Further, tradeoffs between fairness criteria in machine learning have not been addressed with regards to organizations. We advance the research by introducing an organizing framework for selecting and implementing fair algorithms in organizations.

**ACKNOWLEDGEMENTS:** The authors thank United States Agency for International Development (USAID) Grant AID-OAA-A-12-00095 "Appropriate Use of Machine Learning in Developing Country Contexts" awarded to MIT D-Lab which funded authors 2 and 3; the first author thanks John Chambers College of Business and Economics at West Virginia University and all authors thank the Carroll School of Management at Boston College for research support. The first author thanks Research Assistant Mariana Paredes, Research Assistant Peixuan Huang, and Research Assistant Madison Choo (of Boston College). The authors thank Daniel Frey (MIT), Sam Ransbotham (Boston College), Robert Fichman (Boston College), Daniel Brown (Harvard Business School), Aubra Anthony, Shachee Doshi, Craig Jolley, Amy Paul, Maggie Linak (all USAID), Rich Fletcher, Amit Gandhi, Lauren McKown, Kendra Leith, Nancy Adams (all Massachusetts Institute of Technology), and John Deighton (Harvard Business School) for valuable feedback. We also thank the anonymous referees at Academy of Management Annual Meeting 2020, Strategic Management Society Annual Meeting 2020, and Society for Business Ethics Annual Meeting 2020 for valuable feedback.

## Introduction

Following the wide availability of Machine Learning (ML) throughout the last decade, it has been tempting to apply ML algorithms to a variety of business contexts and to automate tedious and costly organizational tasks. However, the powerful advancements of ML in the digital world also carry professional and ethical responsibilities (Leicht-Deobald et al., 2019; Martin, 2018). If organizations implement ML in the absence of a carefully curated training set, awareness of its ethical implications, and human supervision of the tool, neutrality is sacrificed (Mann & O'Neil, 2016). Amazon's now-disbanded ML recruitment program that reviewed job applicants' resumes serves as a clear warning of the dangers of ML for perpetuating discrimination and unfairness (Dastin, 2018). Originally intended to filter through hundreds of resumes to select qualified candidates, Amazon's ML took an unexpected turn when it was discovered to have developed a bias against women. The bias emerged from the data used to train the ML algorithms, data which consisted of actual resumes submitted to Amazon over a ten-year period. Because most job applicants in the data pool were male, the program determined that men were more qualified candidates than women. Specifically, the ML model learned to penalize resumes that included the word "women," such as by downgrading candidates who belonged to all-female extracurricular groups or had graduated from all-women's colleges.

As observed in the Amazon incident, the inner workings of ML algorithms are often poorly understood by organizations, even by those within the ML field who are adept at calibrating and implementing such models. Certain algorithm classes, such as neural networks, are solutions whose intimate processes are inherently hidden and unintuitive (Serra, 2018). We add to the growing corpus of research in business ethics and management that examines how organizations can more readily address the problem of bias in ML, providing pragmatic solutions and guidelines for those seeking to fairly implement these algorithms. By fairness, we broadly refer to global perceptions that decisions and procedures adhere to agreed-upon rules about equitable treatment (Ambrose & Schminke, 2009; Newman et al., 2020).[1] While issues of fairness in ML are newer to our field, the laws protecting against discrimination were set in place decades ago, long before ML existed as a widespread technology. The fact that the penalties in many of the laws are imposed on the individual level, as the laws set penalties *per individual event*, poses high risks for firms using automated processes that may inadvertently discriminate millions of times a second and so subject the companies to very large penalties (Ajunwa, 2020)—similar to the legal risks of data breaches.

This paper seeks to clarify fairness issues as they relate to ML, which are timely because of the growth of automation within organizations, and proposes an organizing framework by which firms may mitigate fairness risks throughout different phases of ML use, from errors in the initial design of ML to improper human oversight after the model is executed. We synthesize research in computer science, information science, management, and behavioral ethics in order to reveal opportunities for organizations to proactively manage ML. In particular, we examine several popular algorithmic metrics designed to improve fairness in organizational settings: fairness through unawareness, demographic parity, equality of opportunity, and equality of odds. We develop a decision tree diagram to help firms more effectively determine which fairness metric to apply given their fairness concerns, data, and situational constraints; in doing so, we extend recent work encouraging greater accountability for firms in the design and development of algorithms (Kroll et al., 2017; Martin, 2018). In addition, we shed light on two forms of human bias, ethical fading and overconfidence bias, that may interfere with ML implementation and suggest potential solutions for more effectively oversight of ML. Without such managerial intervention, ML runs the risk of perpetuating discrimination and bias against marginalized groups. We conclude by discussing the applicability of the framework across various organizational settings, as well as the limitations of the framework, which suggest future research directions.

---

[1]Fairness and justice have historically been treated as interchangeable terms although more recent work has viewed them as distinct concepts (Colquitt & Zipay, 2015; Goldman & Cropanzano, 2015; Greenberg, 2011). Given our broad conceptualization in this paper, we regard fairness and justice as interchangeable.

## A Brief Review of Fairness in ML

ML is the study of automation models which improve with more data. As a field of research, ML is at the intersection of computer science, statistics, linguistics, and mathematics. Only recently did issues of fairness and ethics begin to take a more prominent role in ML scholarship, beginning with the work of Dwork and colleagues (2012). The key objective of ML is to enable predictions which improve as additional data becomes available to the algorithm. While ML tools are promising in their ability to provide substantial increases in organizational efficiency and productivity, there are also significant shortcomings that should not be overlooked, namely those which may negatively impact fairness. To identify these shortcomings, it is important to first understand how ML operates in practice.

ML involves the use of computer algorithms—that is, a set of machine-computable instructions which solve a problem in a finite number of steps (Corman, 2009)—to automatically create models from data. A *classifier* is an algorithm that predicts a categorical outcome. ML algorithms learn from existing labeled datasets where the observed outcomes are labeled for every predictor variable, otherwise known as a *feature* (i.e., the two terms are equivalent and used interchangeably throughout the literature; James et al., 2013). For example, a bank may have data profiles of applicants who filed for a mortgage application as well as the bank's decision for each applicant. This labeled dataset is then split into a *training set*, which is used for the ML model to *train* or *tune* its parameters to fit the data, and a *test set* which leaves out part of the known (labeled) data in order to verify whether the algorithm is producing correct predictions (for reviews, see Marsland, 2015; Teodorescu, 2017). For a binary criterion variable, such as hiring a job candidate or not, the outcome is categorized as follows: true positive (candidate is hired), true negative (candidate is rejected), false positive (algorithm predicts candidate will be hired but candidate is rejected), and false negative (algorithm predicts candidate will be rejected but candidate is hired). These four values are then used to determine the accuracy of the algorithm, which is calculated by the ratio of correct predictions (actual hires, actual rejections) versus the total number of predictions attempts. Accuracy is the primary and often only measure used by computer scientists to evaluate the performance of a ML algorithm and says virtually nothing about the fairness of the model.

Figure 1 shows a stylized illustration of misclassification errors in a hiring situation where the outcome is labeled as "positive" or "negative" across gender. Panel B displays a simple classifier splitting the data from Panel A into predicted positive and predicted negative with the false positives and the false negatives circled. The stylized example represents a typical situation in ML classifiers where a majority of the predicted positive values are actually positive outcomes and a majority of the predicted negative are actually negative outcomes, yet there are some misclassified (such a classifier would have a fairly good accuracy).

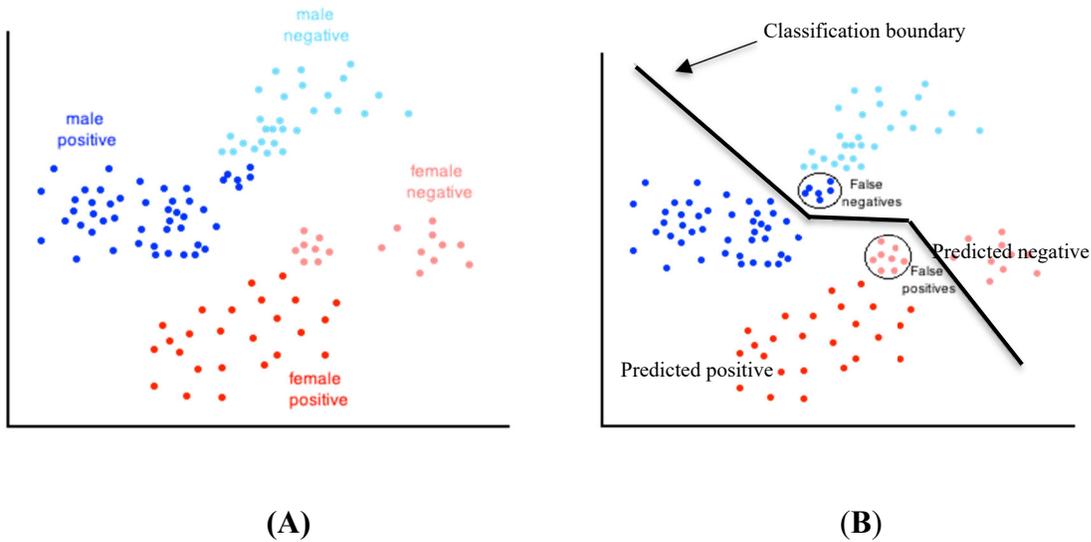

**(A)** **(B)**

**Figure 1.** Example of a dataset with one feature (i.e., gender) and one binary outcome (i.e., hired vs. not hired; Panel A) and of misclassification errors (Panel B).

Unfortunately, misclassification is only the tip of the iceberg and instances of harm and unfair bias abound in the ML literature. O'Neil's (2017) work is one of the best-known collections, notably describing a teacher being unfairly deemed a poor worker by a ML model, subsequently leading to her wrongful dismissal. There is also the widely known work surrounding the recidivism prediction system COMPAS, which was found to be prone to unfair prediction and discriminated based on race (Brennan et al., 2009). Figure 2 displays a synthesis of typical sources of unfairness based on the ML literature in the order they are normally treated when designing an ML algorithm—starting with data collection, continuing with the feature selection, the model selection, and finally issues pertaining to the actual algorithm. We observe that unfair ML comes in many flavors and may have numerous causes. Common pitfalls of ML include generating false positives and false negatives (Bonta et al., 1998; Bonta et al., 2002; Chierichetti et al., 2017; Grgić-Hlača et al., 2018), sampling on small populations (Brennan et al., 2009), hidden correlations in input data (O'Neil, 2017), and oversimplified or distorted models (O'Neil, 2017). Avoiding such traps may be expensive, but it is imperative that organizations recognize and make efforts toward preventing violations of fairness in ML tools.

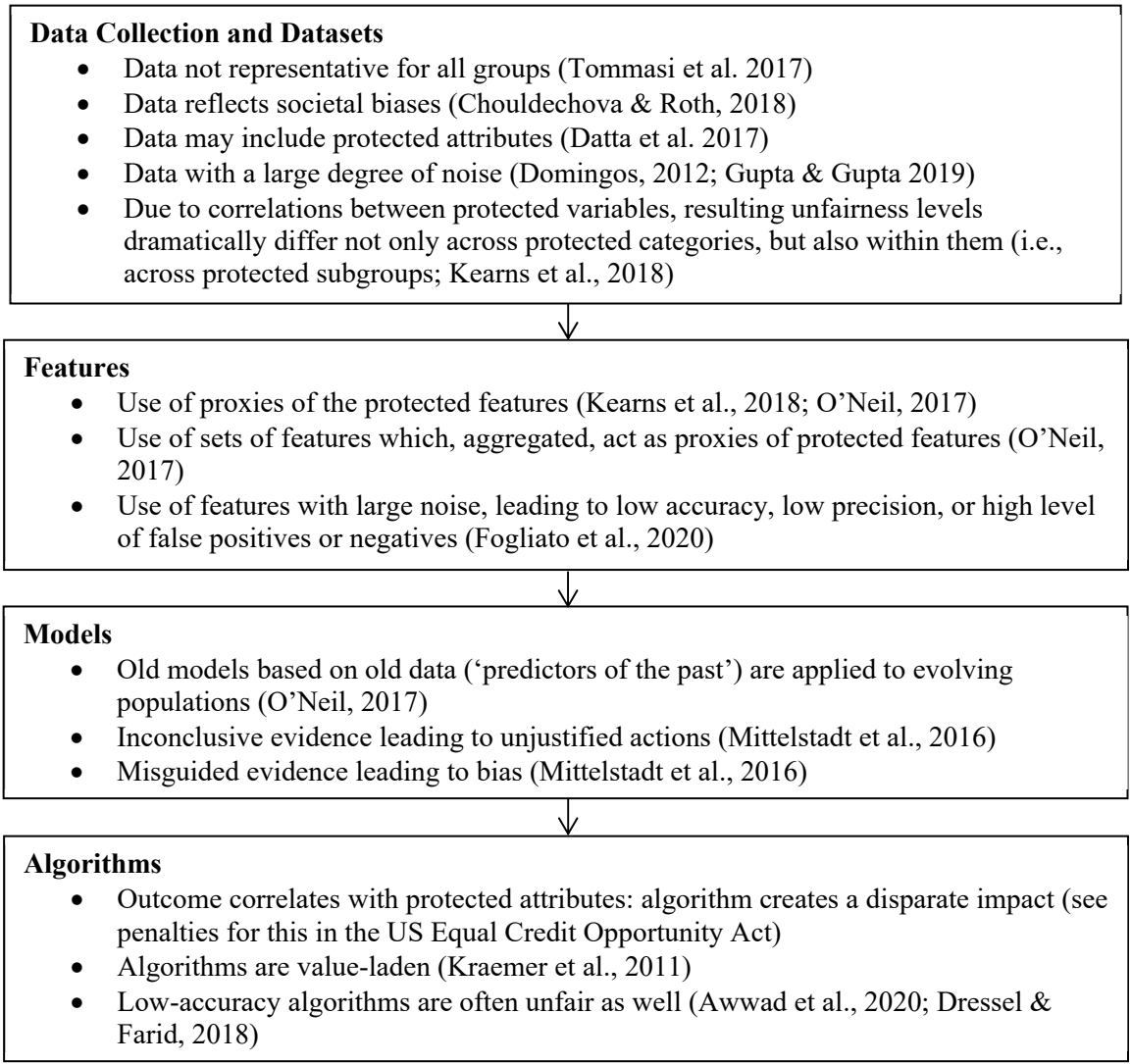

**Figure 2.** A summary of typical sources of unfair ML.

We call special attention to the propensity for ML to engage in stereotyping due to its tendency to make conclusions based on some evidence in the data, however incomplete and socially biased, that provides apparently satisfactory answers with only a partial understanding of the phenomenon or problem. Finding the roots of stereotyping is often difficult, but an essential task, and requires organizations and business leaders to take active steps to incorporate fairness considerations whenever developing and deploying a ML-based system. It is critical for the management discipline to overcome its widespread lack of familiarity with ML and gather a deeper understanding of the technical methods available for applying fairer algorithms. At least as importantly, there is a growing need to develop more structured governance procedures for companies to navigate the complex labyrinth of fairness in ML while also maintaining other priorities, an endeavor which many organizations are not comfortable handling (Agrawal et al., 2018; Osoba & Wesler, 2017).

## A Framework for Fairer ML Within Organizations

In this paper, we seek to hasten progress on the topic of fairness in ML and address the accountability and governance challenges currently plaguing organizations by introducing a conceptual framework for implementing fairer ML (see Figure 3). We integrate research in computer science, information science, management, and behavioral ethics to provide a more comprehensive discussion of the distinct phases of the ML process (i.e., Design, Development, and Post-hoc Model Assessment). Of particular note, we extend the current understanding of the Design and Post-hoc Model Assessment phases by revealing novel opportunities for organizations to more proactively manage fairness. We also contribute to current knowledge of the consequences of ML in the workplace by illuminating two person-driven biases that may prevent ML from being carried out properly.

First, we briefly summarize the key features of the model. The left side of the model indicates that humans and ML mutually shape the ML fairness process, which is depicted by the dotted rectangle in the center of the model and begins with the Design phase (Identify protected attributes, Create an *N*x*N* matrix, Select fairness criterion). It continues with the Model Development phase (Normalize data, Test whether data is representative, Test whether data contains proxies of protected attributes) and ultimately the Post-hoc Model Assessment phase, which includes a human oversight component (Calibrate model, Conduct human oversight of ML decisions, Reassess). The process we describe is also shaped by human-driven ethical bias that, left unchecked, can disrupt efforts to apply fair ML algorithms. The right side of the model indicates that implementing fairer ML is the final outcome; however, it is not necessarily the end of the process. Improper execution of any phase, changes to the data or the external environment (e.g., new variables, adjusted fairness goals, shifts in the organizational environment), or the observation of unintended outcomes can trigger recursion and lead firms to cycle back to earlier phases in the ML fairness process for renewed model configuration. We describe the model and illustrate its applicability using the hiring and recruitment context, although we consider the model to have broad applicability to any organizational setting where ML is considered useful and fairness (or the prevention of unfairness) is an important concern for the organization and/or society.

**Phase One: Design**

To design fairer ML, it is important to conduct a three-part analysis of the training dataset that can identify sources of potential unfairness: whether protected attributes exist in the data, whether predictors correlated with those protected attributes exist in the data, and which fairness criterion is appropriate for the given task.

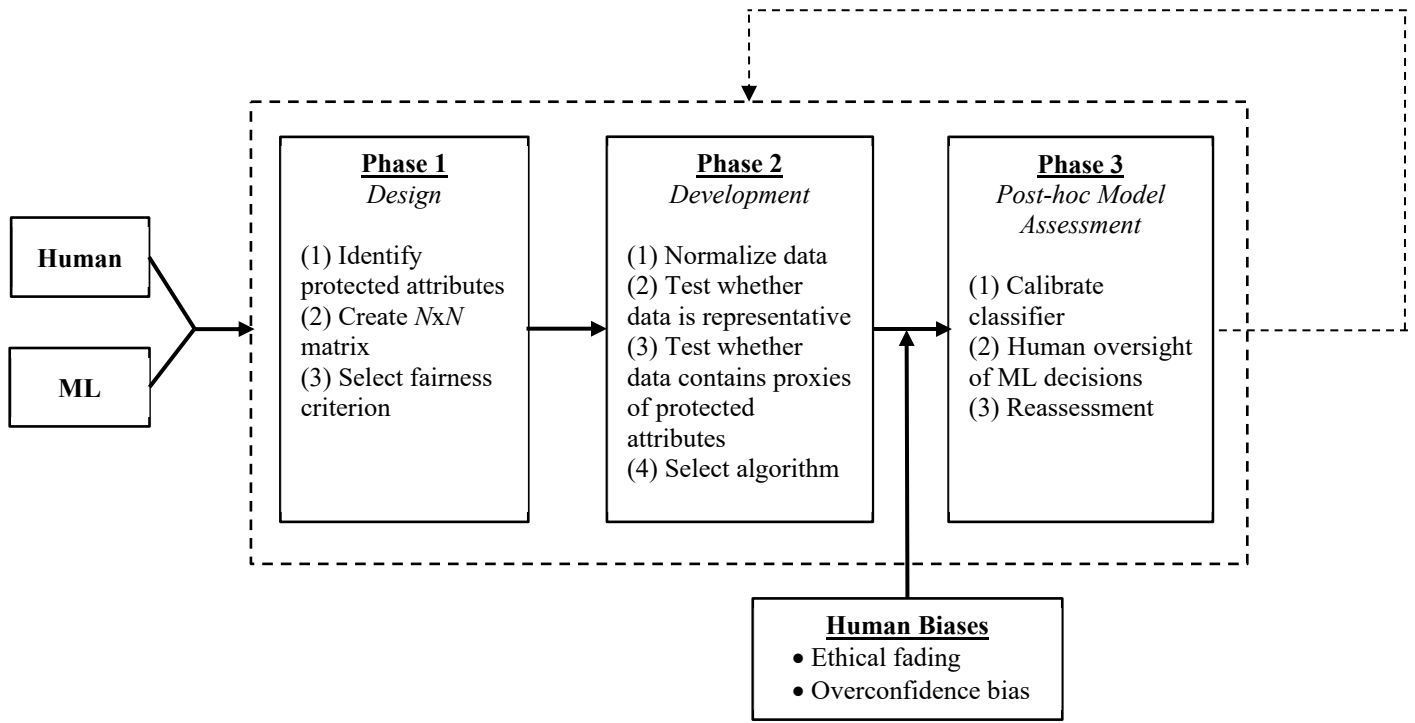

**Figure 3.** Process framework for fairer ML in organizations.

*Identify Protected Attributes*

The first step in the design phase is to determine which features in a training dataset used for learning constitute *protected attributes*, which represent legally protected demographic features such as race, gender, age, sexual orientation, disability status, marital status, ethnicity, national origin, and socioeconomic status. In the United States, these protected attributes are codified into law through equal opportunity in hiring (FEEO), credit lending (ECOA), non-discrimination based on gender or race (Civil Rights Act Title VII 1964), and non-discrimination based on disability (ADA 1990, Rehabilitation Act 1973). Protected attributes might also include other sensitive characteristics beyond an individual's control, such as health data or family links to delinquency (Holder, 2014). If a feature in the dataset represents a protected attribute, it should never be used as a predictor in the model.

### *Create N × N Matrix of Predictors*

A given predictor in the dataset may correlate with one or several protected attributes which may bias a ML model toward negative outcomes for a protected demographic group. If a protected attribute strongly correlates with one or more predictors, the organization should make efforts to remove multicollinearity issues. However, there is a tradeoff between eliminating all predictors correlated with protected attributes and optimizing model performance. Thus, the second step of the design phase is to take measures to avoid falling into a local-optimum pit, where an algorithm is fair according to one protected attribute but at the detriment of another variable. One way to accomplish this is to randomize variable selection–that is, to select at random among similar features those not correlated with protected attributes. Kendall's tau correlation is a metric for correlating any variables regardless of scale or distribution since it is based on each variable entry's rank rather than value (Abdi, 2007). We argue that calculating the tau correlation table is a necessary step in sorting out structural issues in the data to promote fairer prediction.

For example, if a company wishes to select qualified job candidates from a pool of online applicants, it should first evaluate whether its training dataset displays meaningful differences between protected attributes by conducting an *N*x*N* correlation matrix containing all protected attributes and all remaining features in the model, the latter of which include both the explicitly entered features, such as degree/major, university, and work experience, and the myriad features deduced from resumes and other hiring documents. If there are medium to high correlation values (e.g., greater than 0.5) or other forms of dependency between the rows/columns representing protected attributes and those representing features, then those features should be treated with caution at the very least, or simply treated as protected attributes.

*Select Algorithmic Fairness Criterion: A Decision Tree Approach*

Taken together, the initial design steps–identifying protected attributes and determining whether other variables in the dataset correlate with protected attributes–provide a more carefully considered protocol for enhancing fairness in one's data. In the next step of the design phase, we present a decision tree approach for selecting an algorithmic fairness criterion *prior to developing the ML model* (Figure 4). It is important to emphasize that fair algorithm selection is dependent on the characteristics of the data as well as on the developer's understanding of the organization's goals, fairness issues, and situational constraints surrounding the use of ML.

**Fairness Through Unawareness**

The simplest and most commonly applied approach to fairness by organizations is to simply ignore protected attributes: remove them from the model as if they didn't exist in the data (Kusner et al., 2017; Verma & Rubin, 2018). Because differences in data due to different backgrounds are downplayed and not explicit, the potential for bias is thought to be limited. The problem with this approach is that by removing the protected attributes from the data, organizations are not checking whether there are any predictors used by the model which may be highly correlated with protected attributes (i.e., Step 2 in the Design Phase). As a result, the firm may in fact end up with a redundant encoding. For example, in the case of Amazon's flawed hiring algorithm, certain variables from the text of resumes perfectly encoded gender, such as activities or affiliations which were specifically women's clubs or women's colleges, and were picked up by the algorithm (Dastin, 2018). The fact that Amazon had a problem with gender-specific features in resumes implies that removing this protected attribute did not prevent the algorithm from discriminating based on it, making the fairness-through-unawareness approach a less than ideal method. Thus we argue that fairness through unawareness should be applied only when there is strong evidence that the features used by the algorithm do not correlate with the protected attributes, which is seldom, if ever, true.

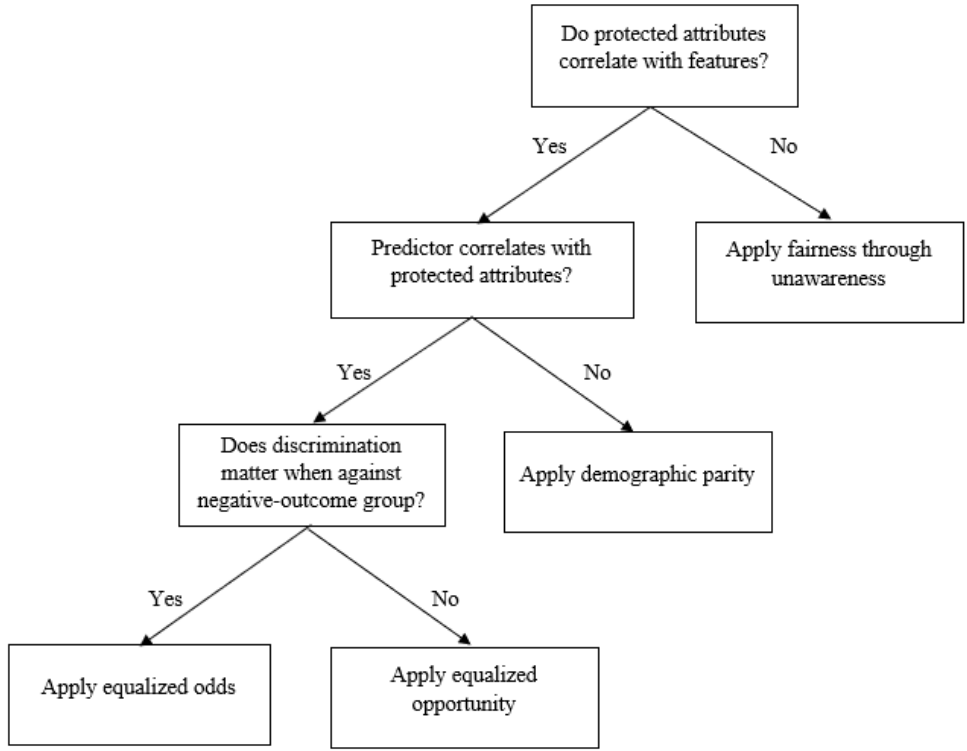

**Figure 3.** Decision tree diagram for selecting an appropriate fairness criterion.

    Alternatively, if a feature correlates with a protected attribute, the organization can decide to remove the feature, although doing so carries the risk of significantly degrading the model's prediction accuracy. To mitigate tradeoffs in model accuracy, another variable or set of variables may need to be included in the data to replace the undesirable feature. For example, when Xerox attempted to create a model to predict churn in prospective call center applicants, it found that commuting distance was highly correlated with poverty (O'Neil, 2016). In this case, the

company could have asked what causes long commuting distances for candidates and collected more data. Removing the variable of commuting from the model could have then translated into adding more features, which in turn would have made it possible to counteract the correlation while potentially increasing the performance of the model. Perhaps the company would have found that employees preferred the countryside, or that two spouses sometimes split the commuting distance between distant workplaces.

The caveat of counteracting multicollinearity is that it is often impractical and sometimes impossible for organizations to collect more data. If an organization cannot precisely determine whether there is a substantial link between a feature and protected attributes, it is safer to assume that there is. If the organization cannot sufficiently mitigate multicollinearity issues, we recommend proceeding with the next node in decision tree to consider other strategies for achieving ML fairness. These fairness criteria take a more proactive approach that addresses data imbalances, but with each deeper intervention comes the risk of over-correcting and forcing equality where it is not to be expected (Dwork et al., 2012).

**Demographic Parity**

Demographic parity is a group fairness criterion where the positive outcome should be reached at the same rate irrespective of the categories of a protected attribute. For instance, if one were to consider gender in the hiring context, the goal would be to separate "strong" from "weak" job candidates while ensuring that men and women are recommended for hiring at the same rate (Kusner et al., 2017). This is a more accountable approach than fairness through unawareness as there is a conscious decision to tune the algorithm so that the positive outcome (i.e., recommended hire) is independent of the protected attribute, thus the protected attribute is not discarded but rather a part of the process of choosing an algorithm that is fair to all categories within a protected attribute.

A major downside of this fairness criterion is that in predicting whether to hire a job candidate, demographic parity has no way to deal with differences between groups other than to assume that the rates of success are equal. In other words, it neglects *individual unfairness*, sacrificing in some cases qualified individuals in order to achieve equality at the *group-level*. For example, the demographic parity approach would not be fair to individuals within a protected group who differ on a non-protected measure, such as analytical ability, programming skills, empathy, communication skills, or any other metric that is relevant for performance for a particular job description. To achieve equal outcome rates across two groups, the organization might hire candidates from group A who are not as qualified as the least-qualified candidate in group B. We would need, however, to hire from the two groups at the same rate: this would be unfair to the qualified but rejected applicants in group B, although on the surface it might appear fair because we would satisfy the condition of parity across the two groups. Therefore, the criterion of strict equality of predicted outcomes across protected categories is difficult to apply in practice. In fact, the "perfect predictor", where all predicted outcomes are equal to the true outcomes is not possible in the case of demographic parity across all dataset scenarios (Dwork et al., 2012; Veale & Binns, 2017).

Demographic parity can also cause certain groups to appear worse than others, leading to a self-fulfilling prophecy (Dwork et al., 2012). Imagine a company that rigorously hires male job applicants at a rate of 35% and indiscriminately hires female applicants at the same rate (Ghassami et al., 2018). Although the acceptance rate in both gender groups is the same, the low effort made to ensure that the best female candidates are chosen—that is, that women are hired who will compete well against male candidates—will likely make the female hires look like poor performers, thus establishing a negative track record for the female group.

It can be argued that demographic parity should increase with time and serve as a long-term goal, though it should not be strictly enforced unless the conditions across protected groups are sufficiently equal to make it a reasonable criterion (e.g., similar qualifications for job position). If the protected groups are too different in their feature values to make demographic parity a reasonable approach, either the nature of the imbalance needs to be explored to determine a proper solution or alternate fairness criteria should be considered.

**Equality of Opportunity and Equalized Odds**

Equality of opportunity measures whether individuals who should qualify for an opportunity have the same likelihood of being deemed as qualified by the ML model regardless of the value of a protected attribute. Like demographic parity, this fairness metric focuses on positive outcomes and additionally ensures that, whatever the value of the protected attribute, the rate of a predicted positive outcome for a qualified individual is equalized (Hardt et al., 2016; Kusner et al., 2017). Thus equality of opportunity is a more targeted version of demographic parity that allows for demographic differences but requires that the unfair/erroneous judgments, or false positive rates, be equitably distributed. In the hiring example, while strong male and female candidates may be of comparable quality, the algorithm could detect a significant difference among the weak candidates (where *weak* corresponds to job performance for the position the firm is hiring for) across gender, such as females in the weak job performance category being more qualified candidates than males.

If implementing equality of opportunity would result in more weak male candidates being hired than weak female candidates, then equality of opportunity is not sufficient and equalized odds should be considered. The equalized odds fairness metric is a stricter version of equality of opportunity in that it adds to the requirements the condition that the true positive rate *and* the false positive rate are equal across categories of the same protected attribute (Hardt et al., 2016). Consider an organizational hiring dataset where participation in college football is positively correlated with particularly strong male job candidates who combine scholastic achievement with athletic excellence. However, weak candidates who played football in college and are male may be considered weaker than other weak candidates because of the time-consuming nature of the sport. This could result in the ML model accepting very weak male candidates at a rate higher than very weak or weak female candidates because of the male candidates' association with very strong job candidates due to the shared attribute of college football. If such confounding variables exist, they may cause equality of opportunity to be insufficient, and equalized odds will need to be enforced. The equalized odds algorithm would ensure that weak female and male job candidates have an equitable shot at being hired by the firm. Although this is a much stronger criterion than equality of opportunity, it is less practical in that there are situations where equalized odds is not true for a particular dataset, algorithm, and protected attribute combination (Hardt et al., 2016; Kusner et al., 2017).

    A key takeaway from our discussion is that ML fairness criteria, as they are currently understood, tend to address one specific fairness failure at a time, which results in a scenario where plugging one leak in algorithmic design results in the emergence or exacerbation of other leaks. Organizations must walk a fine line—they must understand the data distribution, ethical implications, and real-world context from which the data was gathered to enforce just the right amount of fairness in a manner that enhances the model's utility (i.e., accuracy) in the medium- and long-runs. The decision tree diagram we present offers guidance for navigating this tightrope

by providing a step-by-step approach for selecting an appropriate fairness metric. While relying on a single fairness criterion or doctrine is not a complete solution, we argue that it is a more reasonable approach than ignoring fairness criteria altogether or applying every relevant or appealing fairness metric in a non-strategic manner.

**Phase Two: Development**

A significant challenge in addressing specific manifestations of ML unfairness is that the various forms of unfairness are often driven by similar forces, making it unrealistic to resolve each issue in a piecemeal, one-at-a-time manner. Rather, a smart grasp of the correlation landscape is necessary to tackle the problem. Systems moving in sequence, rather than in a noisy random manner, allow the rapid spread of contagion, similar to the spread of a computer virus or an economic crisis. Unfairness in ML propagates due to hidden, poorly understood correlations that were missed in the Design phase. To mitigate this problem, in the development phase, standard steps are taken by developers to prepare a dataset for analysis. These procedures typically involve normalizing the data, testing whether the training dataset is representative of the population, determining whether features that were not correlated with the protected attributes may in fact be proxies for the protected attributes, and choosing an algorithm. Because the development phase has received considerable attention by both scholars and practitioners compared to the other ML phases in our framework, we provide a brief discussion here.

*Normalize Data & Test Whether Data Is Representative*

First, normalization is an expected data processing step (Marsland, 2015). It is important to normalize the training data by bringing all predictors to the same scale (for example, between 0 and 1) as some algorithms are sensitive to scale. Min-max normalization is frequently used, although z-score normalization is expected for some algorithms like support vector machines and neural networks. Next, developers should perform checks to determine whether the data is representative of the population. There are several forms of bias that may skew the representativeness of the data, such as sampling bias or and historical bias (for a review of bias in AI, see Mehrabi et al., 2019). If the data is not representative of all categories within a given protected attribute, it can inhibit the proper implementation of the fairness criterion. Thus it is essential to ensure that the data is representative. There is a science to choosing appropriately representative data that does not create new fairness challenges. More data is not better if the additional data skews the existing dataset further away from the real-world phenomena it represents.

***Test Whether Data Contains Proxies of Protected Attributes***

The third step in the Development phase is more difficult than performing a simple correlation check and involves understanding the phenomenon that generates the training data. It is possible that the dataset may indirectly incorporate protected attributes by including predictors in the model which highly correlate with the protected attributes, effectively revealing what is in the supposedly hidden attributes. For instance, it is now known that in the United States, zip code is a proxy for race (O'Neil, 2017), but this would not be apparent without some background knowledge of the historical trends in housing. The concept of inadvertent discrimination due to societal biases existing in the training data is formalized as "discrimination by proxy" (Datta et al., 2017). In natural language applications, special care needs to be given to the time period of the texts used in training data, as language can directly encode societal biases of a given time period and a developer not paying attention to this may inadvertently create a biased algorithm, (Bolukbasi et al., 2016).

*Select Algorithm*

After successfully testing and debiasing the data, an organization may proceed with choosing an algorithm. Algorithm selection can involve any standard approach deemed appropriate for the organization's needs, such as a tree, a neural network, a support vector machine, or a number of other options (for an overview of algorithms, see Marsland, 2015). The choice of algorithm is usually enacted by the developer who will implement the ML solution for the organization and may involve either the use of the firm's in-house capabilities for custom solutions or the reuse of existing implementations through libraries.

**Phase Three: Post-hoc Model Assessment**

The third phase of our framework involves a combination of machine calibration and human assessment of the ML output derived in the Development phase. During this stage, it is critical for the organization to evaluate the ML model once it has been trained in order to (1) determine whether the algorithm was applied correctly and (2) delineate the algorithm's ethical implications from a business perspective.

*Calibrate Classifier: Posteriori Tests for Fairness*

We begin with calibration of the algorithm using posteriori tests for fairness. Prediction accuracy is important to organizations, especially when the prediction-aided decisions may have fundamental consequences for employees or stakeholders (Mason, 1986), and this accuracy is an essential part of adoption by users; if error rates are low, users are likely to accept the system (Gill, 1995). Tests for the correctness of the algorithm and data should be systematically performed to correct any errors and biases in the data used to train the ML model.

In the organizational hiring scenario, a posteriori false positives (e.g., weak employees who were hired) are generally easy to determine, but not false negatives (e.g., missed hiring opportunities). If the organization wishes to better understand the algorithm's observed rate of false negatives, it might examine the post-denial employment trajectory of the pool of employees who were shortlisted for hiring but ultimately rejected (de Cuerto, 2012), perhaps by using LinkedIn profile data. Testing for false negatives could also be conducted by including a percentage of people 'not recommended for hiring' in the 'recommend for hiring' population via random selection. For instance, if the hiring algorithm is based on a score of 1 to 100 and the minimum threshold for a hire is 90, taking a small random sample of people with scores between, say, 80 and 90 would make it possible to estimate the probability of false negatives. This, in turn, could allow the organization to adjust the algorithm to minimize the number of false negatives and so to be fairer to the job applicants. In addition, the numbers of false negatives and false positives are indicative of data-driven biases. Tests involving computations of the false positives, false negatives, and true positives and true negatives are standard in ML, and it is common practice to run several such measures as part of the calibration of the algorithm (Bella et al., 2010; Marsland, 2015, Pleiss et al., 2017).

### *Human Oversight of ML Decisions*

Because the sophisticated techniques involved in ML design and development run the risk of introducing systematic unfairness into organizational processes, it is vital for individuals to constantly monitor and revise the quality of the ML model. Drawing from research in management and behavioral ethics, we outline promising solutions for organizations to help safeguard against breakdowns in fairness.

A common line of defense against inaccurate performance and unintended consequences in organizations is the formation of an independent oversight committee. Oversight (also referred to as audit) committees represent a form of corporate governance in which boards of qualified organizational members develop and maintain responsible business practices while taking into account the interests of the communities within which they operate (Nolan & McFarland, 2005; Sonnefeld, 2002). In ML settings, oversight committees should play a strategic role in supervising algorithmic fairness processes and preparing advice regarding the quality and appropriateness of the ML algorithm's output. Indeed, empirical research indicates that effective oversight committees are positively associated with more accurate task performance (Karamanou & Vafeas, 2005). Individuals selected to serve on such committees, including the chairperson, should have a strong background in the domain addressed by the algorithm (in case of employment/hiring, expertise would be measured by previous experience in hiring) and possess sophisticated knowledge of ML principles and techniques. Committee members without proper expertise in these two areas, though well-intentioned, may be less precise in their supervision, or worse, can derail attempts at fairness by promoting the use of flawed algorithms. Generally speaking, oversight committees should review their composition periodically to confirm that members have the knowledge and experience they need to be effective and should be reviewed at least annually. If the organization cannot find employees with adequate knowledge in these two areas, it is recommended that the committee be composed of cross-functional members who are experts in at least one domain.

Before the audit committee meets, the company should first prescribe the committee's recurring responsibilities, such as explicitly evaluating whether the company's ethical values and legal principles are recognized in the ML model's output. If ethical issues are flagged, it is the audit committee's responsibility to recommend priority improvements for calibrating the algorithm. For instance, if in this oversight phase a company discovered a bias in its hiring algorithm against female applicants (e.g., by checking a sample of the resumes and predicted outcomes manually by the oversight committee), the bias could be addressed and fixed before the hiring algorithm is released throughout the entire organization.

On the more technical side, human oversight of ML algorithms should involve probing the algorithm for potential unfairness against subgroups of the protected attribute groups presented. For example, if fairness is to be enforced for gender and racial groups, then gender-racial groups, such as women who are also minorities, must also be protected against unfairness. Often the trouble with ML fairness criteria is that they bluntly enforce specific definitions of fairness that do not always work together symbiotically, especially when released into the "wild". This remains an area of active research (Kearns et al. 2018) and continues to pose implementation challenges. Because searching every combination of subgroups and comparing them against one another is prohibitively expensive, a more strategic approach is necessary to explore the data co-relationship landscape. We argue that organizations should consistently monitor correlations between all of the variables, particularly noting sharp negative and positive correlations, and take efforts to translate such mathematical relationships back into their socioeconomic meaning.

We provide one technique for quickly determining through a visualization whether subgroup unfairness has occurred post hoc. This approach helps to address the issue of hidden information in correlations between the features and attributes, including information that—in principle—ought not be decisive in the outcome. The problem of optimizing on a protected attribute only to find out that adding a second protected attribute creates an unfair outcome is widely described as the driving problem behind the challenge of fairness gerrymandering (Binns, 2018), yet multivariate quantitative techniques and network science methods remain under-utilized in the area of ML fairness. While developers may easily pick up on correlations between individual variables, more complex multivariate correlations and relationships are less obvious (but just as obvious to an algorithmic classifier). We argue that testing subgroups is essential for firms to ensure that a classifier is indeed fair. This can be accomplished through the use of visualizations, which aid a human overseer in observing whether a classifier is clustering the data based on a protected attribute. We create a stylized example in Figure 4 which shows two unfair algorithms. In Panel A, discrimination occurs when the linear classifier cuts the data into positive and negative outcomes based on gender. Panel B displays discrimination when a linear classifier cuts the data into positive and negative outcomes by marital status.

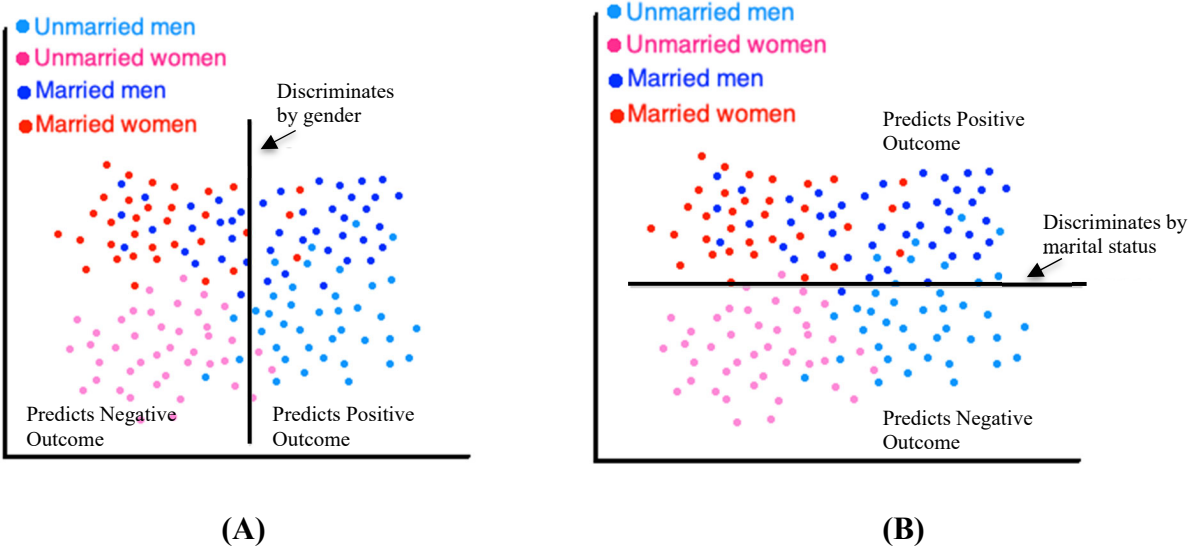

**(A)** **(B)**

**Figure 4.** Examples of unfair algorithms based on gender (Panel A) and marital status (Panel B).

Picture a hiring scenario where the pool of job candidates includes married and unmarried people of both genders with the gender differences being somewhat larger among unmarried people. Again, there is a binary outcome (i.e., hired vs. not hired). It is assumed that married men are unequal to (or, in the plot, separate from) unmarried women but not vice versa, reflecting the societal bias that inequality in economic status within a married couple is likely to favor the husband. In Panel A, the married male cluster is less gender-mixed than the married female cluster. Having four fairly distinct subgroup clusters, a simple linear separator, such as a linear support vector machine, will likely choose one bias or another. Panel B portrays a similar situation of unfair separation except this time the algorithm discriminates by marital status.

In contrast, Figure 5 illustrates a separation which is fairer for both gender and marital status because it does not separate outcomes based on either protected attribute. This can be achieved with a standard algorithm like a support vector machine with a radial basis function kernel. Specifically, developers would create a separation in the data that 1) takes into account the shape of the data and the interaction between gender and marital status, and 2) appears randomly overlapped over the two attributes so as not to distinguish outcome based on the protected attributes. This approach allows the outcome of interest (e.g., hired vs. not hired) to be independent of both gender and marital status. We encourage developers to incorporate visualization tools that review the algorithm after it is run—and, if possible, in the Design phase as well. In doing so, developers can more quickly spot and address cases of egregious unfairness, as shown in our stylized examples.

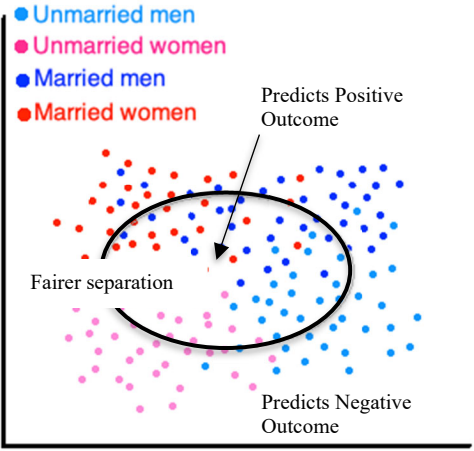

**Figure 5.** A fairer algorithm: the separation is not based on either gender or marital status.

After initial testing of the ML model, the oversight committee should lead its implementation in the organization. It is important that the committee meet frequently to remain informed and knowledgeable about ML issues and how well the company's existing technological systems can respond to them (Abbott, Parker, and Peters 2004; Xie, Davidson, and DaDalt 2003; Nolan and McFarlan 2005). Should an issue arise, the oversight committee should assist ML specialists in resolving the problem in a timely fashion. To prevent substantial speed bumps or mishaps that may arise in the real world, it is advisable to start with a small test case and slowly expand the scope and access of the ML model in the organization. One approach could be to "sandbox" the algorithm and create milestones under observation by the oversight committee prior to a wider rollout within the organization. The committee members should test their ML model with a wide range of unexpected variables to ensure they are considering every possible outcome of the data. Popular approaches to text classification such as word embeddings are a good example of unexpected associations that promote bias (Bolukbasi et al. 2016).

Importantly, it is not immediately clear what effects ML algorithms will have on an organization's outcomes. While it would be tempting to simply deploy the ML solution and, after some performance metrics are achieved, to forget about it, this approach could lead to damaging results for the organization. For example, regardless of whether discrimination is intentional, failure to detect and correct it still leads to liabilities for the organization (see Agrawal et al. 2018). Instead, the organization and particularly the oversight committee members should maintain close awareness of the ML system, noting how it is responding to new data input, how it is changing over time, and of course, whether any significant anomalies have been detected, the latter of which should prompt the deactivation of the system as well as our conceptual model's reassessment step to determine the cause for the anomalies. Thus, the role of the oversight committee is integral and ongoing, requiring thorough assessments of the ML system. The success of an audit committee will depend on how forthcoming the company is regarding sharing its algorithm and data outside of the development team directly responsible for the ML algorithm implementation. Firms may be reluctant to share with outside experts as not to lose competitive advantage to other firms. While we'd argue that a neutral outside audit team would be the approach most likely to achieve fairness, even having an external auditor within the firm but not in the direct control of the unit developing the machine learning algorithm would be beneficial to ensure principles of fairness are applied.

*Reassessment*

The accuracy of an algorithm can change over time with new tendencies in the data. In the case of hiring, such new tendencies might include changes in skills of the applicant population, changes in the needs of the company, and broader socio-economic changes in the applicant pool that may affect the associations made by the algorithm. A periodic reassessment of the fairness of the algorithm is therefore recommended. This reassessment would be a recursive step, transitioning from applying the algorithm in production back to earlier phases of ML.

*Moderator: Human Biases*

It is well known in the behavioral ethics and management literatures that human biases may impair decision making procedures and outcomes, independent of the decision itself (for a detailed review, see Bazerman & Tenbrunsel, 2012). Subconscious biases pose an ever-present threat to organizational fairness, including contexts involving ML. In the present article, we argue that biased perceptions of ML may influence whether the Post-hoc Model Assessment phase will be properly executed by human overseers. While a comprehensive account of human bias is beyond the scope of this paper, we shed light on two biases that we believe carry a high risk of nudging individuals away from issues of fairness and ethical values: ethical fading and overconfidence.

Why would ML that is designed and developed to promote fairness cause people to make unfair decisions without realizing that they have deviated from their values? To answer this question, we draw on a key concept in the behavioral ethics literature known as ethical fading. Ethical fading is defined as the "process when the moral colors of an ethical decision fade into leached hues that are void of moral implications" (Tenbrunsel & Messick 2003, p. 224). In organizational hiring, ethical fading may occur when managers simply do not "see" ethics as being a part of the relevant decision criteria; thus, the organization's values are not reflected in hiring outcomes. Importantly, Tenbrunsel and Messick point out that situational cues shape the decision frame through which a situation is viewed, which in turn influences the likelihood of unethical outcomes. With this research in mind, the present paper cautions that organizations' increasing reliance on ML may disproportionately affect the extent to which moral concerns are salient in decision making processes, ultimately threatening the ethicality of decisions that were previously made by employees.

Indeed, empirical evidence in the behavioral ethics literature suggests that heavy use of technology does not bode well for the maintenance of high ethical standards. Research shows that individuals who rely on non-face-to-face (NFTF) forms of interaction rather than face-to-face (FTF) forms tend to focus more on material outcomes (Hoffman et al. 1996), feel less personally accountable for their actions (Diener et al. 1976), and have less empathic concern for others (Woltin et al. 2011)—all precursors to biased decisions. When organizations use advanced technologies that distance individuals from other people, such as ML, it may cause awareness of ethical implications to fade from human decision making. Through this ethical fading process, organizational members may end up making choices they would otherwise condemn if they were consciously aware of their implications.

We argue that ethical fading plays a critical role when organizations decide whether to move forward with the output provided by a ML algorithm. Developers or managers might naively assume that the ethical component of decision making is adequately handled by the fairness algorithm and no longer requires input from human beings (Leicht-Deobald et al., 2019; Martin, 2018). In other words, ML may "fade" the ethical implications of human actions, leading individuals to henceforth be driven less by values and ethical issues than by feasibility and pragmatic concerns.

In a similar vein, ML may also produce biases of overconfidence. Generally speaking, overconfidence is the subjective confidence that one's judgments are reliably better than the objective accuracy of those judgments (Moore & Healy, 2008). Overconfidence is rooted in an egocentric tendency to overemphasize one's own point of view. Consider the case of ML models used by educational institutions to evaluate teacher effectiveness. As discussed by O'Neil (2017), principals may become overly confident in the accuracy and usefulness of a ML model's decisions—especially if earlier steps were taken by developers to remove sources of unfairness in the training data—leading them to misinterpret facts and shortcomings of the model in order to maintain biased beliefs that the model is fair.

If overconfidence is coupled with ethical fading, it can serve as a recipe for disaster, derailing even the best-laid plans for improving fairness and potentially resulting in costly organizational fines for violating laws and regulations. Thus individuals must not absolve themselves of responsibility for correcting their bias about ML, which begins with awareness. This is easier said than done; for example, reports by popular press have presented conflicting views that ML algorithms are fairer than human beings (Miller, 2018). Although there are instances where the correct use of ML can substantially improve fairness (indeed that is the goal), ML can just as easily obscure ethical standards and conjure biases that, without management intervention, may lead individuals to make sub-optimal decisions.

## Moving Forward: A Summary and New Horizons

This paper provides a comprehensive review of the dangers involved in applying ML analytics in organizations, integrating research in computer science, information science, management, and behavioral ethics. In the spirit of interdisciplinary scholarship, we present a theoretical framework for improving fairness in ML that offers both breadth and depth—our model takes into account the complex interplay between human and machine, technology and organizations, flexibility and rigidity, as well as inherent tensions among fairness and utility. We argue that a systematic approach to model design, model development, and post-hoc assessment and calibration serves as the basis of fairer use of ML and analytics in the business domain. A lack of correctness in ML implementation and the resulting unfairness towards a protected group can be enormously detrimental to firms that develop or use analytics. Such errors expose the organization to huge liabilities, loss of market share, increased employee turnover, and brand reputation costs, in addition to moral violations that transcend monetary concerns. The consequences are even more dramatic when analytics are used on a state-wide scale. They can induce poverty and discrimination, sow social disagreement and unrest, and wreck economic structures.

To avoid such responsibilities and potential costs, an organization may attempt to defer responsibility to the software developer or a third party, however, reputational costs, financial liabilities, and moral failures fall on both the user (i.e., the organization) and the implementer (i.e., the developer). As a result, this work promotes the viewpoint that organizations must proactively implement structured fairness frameworks in their algorithm design, such as the framework proposed in the present article, and delegate appropriate organizational resources and oversight to enforce human accountability in the medium- and long-term.

The ideas and solutions presented in this paper are subject to limitations that suggest future research directions. A key assumption of our model is that organizations have clearly specified goals and values, adequate data collection, resources dedicated to mitigating fairness concerns, and appropriate knowledge of situational constraints. Realistically, it is unlikely that firms will fulfill all of these dimensions, at least at the same time, which will introduce further ambiguity and complexity in the ML process than we have specified. On a related note, certain firm characteristics and behavior, such as problematic market capitalization and a history of discrimination incidents, may raise a red flag to stakeholders and ethical watchdogs, increasing the number of obstacles on the pathway to achieving fairer ML. It is important for business ethics scholars to provide guidance for firms to climb out of such difficult situations, either by offering pragmatic, empirically-based advice or by providing a philosophical understanding of the meaning and consequences that such situations have for organizations and their stakeholders. Lastly, we encourage future research to explore how subconscious biases, beyond ethical fading and overconfidence bias, manifest in the ML context and how they might be minimized. For example, it is likely that status quo bias, defined as a preference to maintain one's current state of affairs rather than change the environment (Samuelson & Zeckhauser, 1988) leads developers to prefer current industry practices toward ML fairness, even though many are subpar in satisfying fairness and underemphasize the responsibility and role of organizations. More specifically, status quo bias may perpetuate the use of the fairness through unawareness metric simply because it is the default in the field. Overall, we are optimistic that the study of ML fairness in organizations is a fruitful area of research for both ethics and management scholars. Our hope is that this paper will accelerate theoretical development and empirical studies that address the question of how to institute fairer ML in organizational procedures and applications.